\documentclass[11pt]{article}
\usepackage[pdftex]{graphicx}
\usepackage{graphicx}
\usepackage{amsmath}
\usepackage{amsfonts}
\usepackage{amssymb}
\usepackage{float}%

\usepackage{enumerate}  

\usepackage{caption}  
\usepackage{subcaption}  

\usepackage{multicol}  
\newcommand{\w}{\omega}

\begin{document}

\begin{center}

{\LARGE{Identifying Resonant Poles by Visual Inspection of Pole-Zero Plots}}
\footnote{Submitted to {\it IEEE Signal Processing Letters} 13 March, 2022; revised 31 May 2022; accepted 12 Oct. 2022.}\\
\bigskip
{\large Amro El-Jaroudi and  Patrick  Loughlin}\\
University of Pittsburgh\\
\end{center}

\bigskip

\noindent{{ \emph{Abstract} -- We derive the $s$- and $z$-plane pole regions for continuous-time and discrete-time LTI systems to yield resonance.  In this way, resonance can be identified by visual inspection of the pole-zero plot, without the need for calculations.
  }}

\vspace{0.15in} 


\section{Introduction}
Resonance is a phenomenon arising in many areas of science and engineering, including electrical circuits, control systems, optics, structural mechanics, among others.  Students in engineering disciplines such as biomedical, civil, electrical, and mechanical  learn that resonance occurs when the damping coefficient of the continuous-time linear time-invariant (LTI) system transfer function is less than $1/\sqrt{2}$.   In contrast to such direct calculations, 
pole-zero diagrams enable one to determine many properties of a system by simple visual inspection, such as whether the system is stable, or low-pass {\it v.} band-pass {\it v.} high-pass, or whether the impulse response will oscillate.  
Identifying resonance from a pole-zero diagram, however, does not seem to be commonly taught as it does not appear in prominent texts in the field (e.g., \cite{Opp}--\cite{Siebert}). (The topic appears in a 1964 monograph on pole-zero patterns for continuous-time systems \cite{Angelo}).
Rather,  the common way that students are taught to determine resonance is to calculate the damping coefficient (for continuous-time systems).   

The primary contribution of this letter is in the teaching of resonance to engineering students, in that we show how resonance may be identified by visual inspection of pole-zero diagrams for both continuous-time and discrete-time LTI systems.  We first derive the region for resonance in the $s$-plane for continuous-time systems and then extend these results to discrete-time systems in two ways:  (1) solving for and plotting the conditions for resonance directly in terms of the $z$-domain system transfer function; and (2) mapping the $s$-plane resonance boundary to the $z$-plane via three common mappings (namely, bilinear, backward, and impulse invariant).   The results obtained extend the power of pole-zero diagrams by allowing for the determination of resonance by visual inspection, without the need to perform calculations.
We also define a discrete-time ``damping coefficient''  and give the numerical constraint for resonance.

\section{Continuous-time LTI systems}\label{CT}
Since resonance requires at least two complex-conjugate poles, consider the underdamped, stable, causal 2nd-order system with poles
$p_1\,=\,p_2^\ast \,=\,-\sigma_0+j\w_0$ and transfer function
\begin{equation}
H(s) = \frac{1}{(s-p_1)(s-p_2)} =\frac{1}{s^2 + 2\zeta\w_n s + \w_n^2} \label{Hs}
\end{equation}
where $-\sigma_0 < 0$ and $\w_0 > 0$ are the real and imaginary parts of the pole $p_1$, 
$\w_n>0$ is the natural frequency, and $0<\zeta<1$ is the damping coefficient. 
Equating the denominators, one obtains
\begin{align} \label{natural-frequency-and-damping}
\sigma_0 &= \zeta \w_n, \qquad \w_0\,\,=\,\,\w_n\sqrt{1-\zeta^2}
\end{align}
As is well known, this system will exhibit resonance ({\it i.e.}, $|H(\w)|>|H(0)|$ for $\w>0$) when  $\zeta < 1/\sqrt{2}$ (or equivalently when the quality factor $Q=1/(2\zeta)>1/\sqrt{2}$).
To find the region of the $s$-plane defined by this constraint, we use the relationships in Eq. \eqref{natural-frequency-and-damping} to obtain the ratio $\left(\w_0/\sigma_0\right)^2$, rearrange to solve for $\zeta^2$, then impose $\zeta < 1/\sqrt{2}$; doing so yields
\begin{equation}
\zeta^2 \,=\, \frac{\sigma_0^2}{\sigma_0^2+\w_0^2} \,<\, \frac{1}{2}\, \, \rightarrow \,\, \left|\frac{\w_0}{\sigma_0}\right|\,> \,1
\end{equation}
The latter result is intuitively satisfying: resonance requires that the magnitude of the imaginary part of the pole be greater than that of the real part of the pole.
Hence, resonant poles can be readily identified by visual inspection of the pole-zero plot:  complex-conjugate poles lying in the left-half $s$-plane region defined by $|\frac{\w}{\sigma}|>1$ will give rise to resonance (shaded blue in Fig. \ref{fig1}a).  The resonant peak occurs at the frequency $\w=\w_r=  \w_n\sqrt{1-2\zeta^2}=\sqrt{\w_0^2 - \sigma_0^2}$, and $|H(\w)|>|H(0)|$ over the frequency range $0<\w<\sqrt{2}\,\w_r$ (see Appendix A).

\section{Discrete-time LTI systems}
Similar to the continuous-time case, we consider the causal, stable 
2nd-order discrete-time LTI system with  complex conjugate
poles $z_{1}=z_2^\ast=Ae^{j\Omega_{0}}$, with magnitude
$0<A<1$ and angle (normalized frequency) $0<\Omega_{0}<\pi$; the transfer function is
\begin{eqnarray}
H(z) & = & \frac{1}{\left(1-z_{1}z^{-1}\right)\left(1-z_{2}z^{-1}\right)}\notag \\
&=& \frac{1}{\left(1-2\frac{A^{2}}{1+A^{2}}\zeta_{z}z^{-1}+A^{2}z^{-2}\right)}\label{Hz}
\end{eqnarray}
where we have defined a damping coefficient $\zeta_z$ for discrete-time LTI systems as
\begin{equation}
\zeta_z=\frac{1+A^{2}}{A}\cos\Omega_0\label{dampz}
\end{equation}
Unlike its continuous-time counterpart, $\zeta_z$ can be positive or negative, with the sign indicating on which side of the imaginary axis the poles reside.  However, as we see below, it is the magnitude of $\zeta_z$ that serves as the quantifier of resonance.

\begin{figure}[t]  
     \centering
     \begin{subfigure}[b]{0.35\textwidth}
         \centering
         \includegraphics[width=\textwidth]{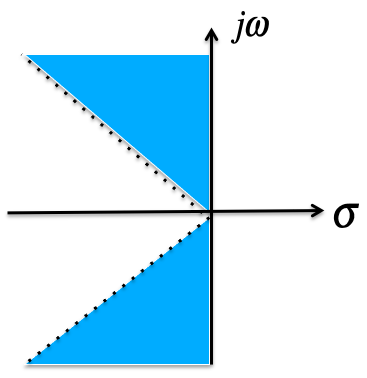}
         \caption{}
         \label{fig:b}
     \end{subfigure}
     \hfill
     \begin{subfigure}[b]{0.37\textwidth}
         \centering
         \includegraphics[width=\textwidth]{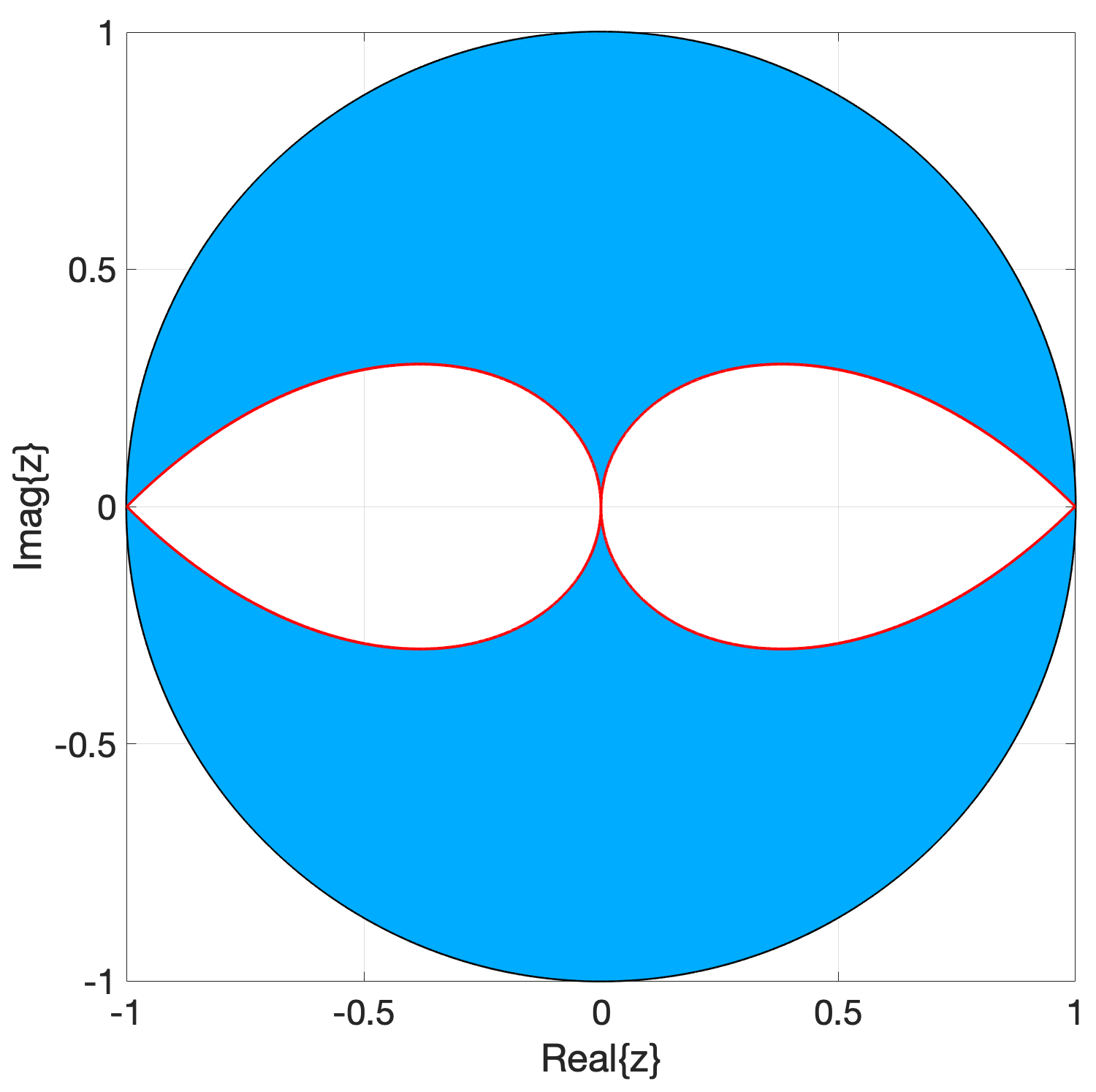}
         \caption{}
         \label{fig:c}
     \end{subfigure}
     \hfill
\caption{\small{Complex-conjugate poles in the shaded region will give rise to resonance. (a) $s$-plane.  (b) $z$-plane.}}       \label{fig1}
\end{figure}

When a resonance exists, the peak magnitude occurs at the frequency  $\Omega = \Omega_{r}$, where  \cite{Proakis}
\begin{equation}
\Omega_{r}=\cos^{-1}\left[\frac{1+A^{2}}{2A}\cos\Omega_{0}\right]=\cos^{-1}\left[\frac{\zeta_z}{2}\right]\label{eq:sol2}
\end{equation}
For a resonant peak, the magnitude of the argument of the inverse cosine must be less than $1$, {\it i.e.,} 
\begin{equation}
\left|\zeta_z\right|<2 \label{damping-z-constraint}
\end{equation}
resulting in the constraint (see Appendix B)
\begin{equation}
\frac{1-\left|\sin\Omega_{0}\right|}{\left|\cos\Omega_{0}\right|}<A<1
\end{equation}
This inequality describes a region in the $z$-plane bounded on the outside by the unit
circle and inside by the curve
\begin{equation}
z=\frac{1-\left|\sin\Omega_{0}\right|}{\left|\cos\Omega_{0}\right|}e^{j\Omega_{0}}\label{eq:boundary}
\end{equation}
Poles within this region (shaded blue in Fig. \ref{fig1}b) will
create resonance in the frequency response. 
Note that for $\Omega_{0}$ near $0$ or $\pi$, $A$ approaches 1,
{\it i.e.}, the pole would have to be very close to the unit circle to create
a resonance. This situation is analogous to the $s$-domain condition where $\omega_0 \rightarrow 0$, for which a resonant pole must be very close to the $j\omega$ axis.  
For $\Omega_{0}=\pm \frac{\pi}{2}$, there is no restriction
on $A$, {\it i.e.}, the pole can be very close to the origin and still
create a resonance. (There is no direct analogy in the $s$-domain.)
Both of these results are intuitively appealing
and consistent with the geometric interpretation of the frequency response 
magnitude utilizing the distance between the location
of the poles and the unit circle.

In addition to the resonant peak occurring at $\Omega_r$, poles in the shaded region
in Fig. \ref{fig1}b 
will result in $|H(\Omega)| > \max\left\{ |H(0)|, |H(\pi)|\right\}$ for frequencies
\begin{equation}
0 \, <\Omega< \, \cos^{-1}\left[\zeta_z-1\right] \label{eq:rangew1}
\end{equation}
for $0<\zeta_z<2$ ({\it i.e.}, resonant poles in the 1st \& 4th quadrants), and 
\begin{equation}
\cos^{-1}\left[\zeta_z+1\right] \, <\Omega< \, \pi \label{eq:rangew2}
\end{equation}
for $-2<\zeta_z<0$ ({\it i.e.}, resonant poles in the 2nd \& 3rd quadrants).
For $\zeta_z=0$, the range is $0<\Omega<\pi$.

\section{Mapping from $s$ to  $z$}
We ask: could we have found the region in
the $z$-plane by mapping the boundary line from the $s$-plane?
We examine three common mappings ($T$ is the sampling period): 
(1) the impulse invariance method,  $z=e^{sT}$; 
(2) the backward difference method, $z=\frac{1}{1-sT}$;
(3) the bilinear transformation, $z=\frac{1+0.5sT}{1-0.5sT}. \,$
Figure 2 shows the boundary line produced for each mapping as well as the exact boundary line derived above. The resonant region predicted by each mapping would be the space between its boundary line and the unit circle.
None of the mapped lines
match the true boundary exactly, although each one approximates the region to
varying degrees for small values of $\Omega_0$ (poles in the 1st and 4th quadrants). In addition, the bilinear mapping
is the closest approximation over all $\Omega_0$, and provides a correct but conservative region, in that all poles identified as resonant are indeed resonant.

\begin{figure}[h]
\begin{center}
\scalebox{0.15}{\includegraphics{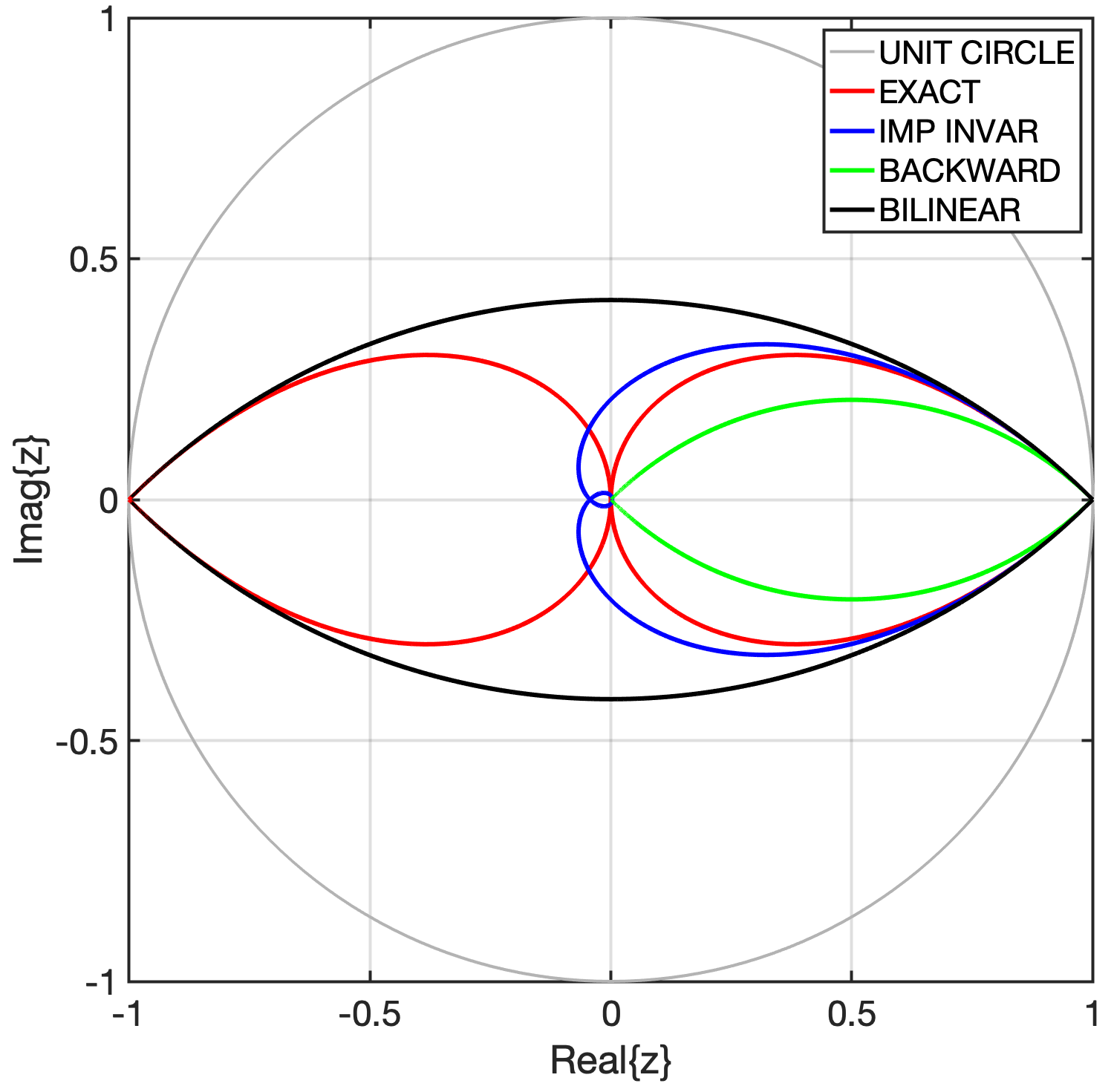}}
\end{center}
\caption{\small{$z$-domain resonance boundary (red) compared to mapping the $s$-domain boundary to the $z$-domain via the impulse invariance method (blue), the backward difference method (green), and the bilinear transformation (black).}}
\label{fig2}%
\end{figure}

\section{Conclusion}
We derived the regions in the $s$- and $z$-planes for which complex-conjugate poles within those regions produce resonance for causal, stable LTI systems (Fig. \ref{fig1}).  As such, resonance can be identified by visual inspection of the pole-zero diagram, without the need to calculate the damping coefficient.  We also showed how the $s$-domain boundary for resonance maps to the $z$-plane using common $s$-to-$z$ mappings (Fig. \ref{fig2}).  Finally, we defined a ``damping coefficient'' for discrete-time systems (Eq. \eqref{dampz}) and gave the condition for resonance (Eq. \eqref{damping-z-constraint}).

\section*{\textsc{Appendix A}}
For pedagogical purposes for the interested reader, we give a derivation for the conditions for resonance to occur for the stable underdamped 2nd-order LTI system with transfer function given by Eq. \eqref{Hs}.
The magnitude of the frequency response is
\begin{align} \label{s-mag}
|H(\w)| &= \frac{1}{\sqrt{(\,\sigma_0^2+(\w-\w_0)^2\,)\,(\,\sigma_0^2+(\w+\w_0)^2\,)}}
\end{align}
We seek the conditions for which this system exhibits {resonance}, {\it i.e.} $|H(\w)| > |H(0)|$. Equivalently, from Eq. \eqref{s-mag}, we seek the frequencies $\omega$ for which
\begin{align} \label{main_constraint_eq}
(\,\sigma_0^2+(\w-\w_0)^2\,)\,(\,\sigma_0^2+(\w+\w_0)^2\,) & < \,(\,\sigma_0^2+\w_0^2\,)^2
\end{align}
Expressing $\w$ as $\w=r\w_0$, where $r>0$, and dividing through by $\w_0^2$ 
we have
\begin{align} 
(\,p^2+(r-1)^2\,)(\,p^2+(r+1)^2\,) & < (\,p^2+1\,)^2
\end{align}
where $p^2=\sigma_0^2/\w_0^2$.  Multiplying out both sides and simplifying yields
$ r^2\left(2p^2+r^2-2 \right) < 0.$
Since $r^2>0$, we must have
$2p^2+r^2-2  < 0$. 
 Substituting for $p^2$ and $r$ and solving for $\w$ yields
\begin{align}
\w < \sqrt{2}\,\sqrt{\w_0^2 - \sigma_0^2} 
\end{align}
Since the frequency $\w$ has to be real and nonnegative, the solution is
\begin{align}
|H(\w)| \,>\, |H(0)|,\,\, \,\,\, &0<\w<\sqrt{2}\,\sqrt{\w_0^2 - \sigma_0^2}, \notag \\
&\hbox{when }\,\frac{\w_0}{\sigma_0}>1
\end{align}
See \cite{Angelo} for an alternate derivation.

\section*{\textsc{Appendix B}}

In this appendix, we derive the conditions for resonance for the discrete-time system of Eq. \eqref{Hz}. The magnitude-squared frequency response is
$\left|H(\Omega)\right|^{2}  = 1/{D\left(\Omega\right)}$, where 
\begin{align}
D(\Omega)  &= 
\left(1+A^{2}\right)^{2}-4A\left(1+A^{2}\right)\left[\cos\Omega\cos\Omega_{0}\right] \notag \\ 
&+4A^{2}\left[\cos^{2}\Omega-\sin^{2}\Omega_{0}\right]\label{eq:defin}
\end{align}
We seek the range of frequencies $0<\Omega<\pi$ for which $|H(\Omega)|^2$ is greater than
$|H(0)|^2$ and $|H(\pi)|^2$, or equivalently
$D\left(\Omega\right)<\left\{D(0), \, D\left(\pi\right)\right\}$.
Using (\ref{eq:defin}) and solving these inequalities, we find
\begin{equation}
\zeta_z-1 <\cos\Omega< \zeta_z+1\label{eq:range}
\end{equation}
where $\zeta_z$ is the damping coefficient defined in (\ref{dampz}).
To obtain the range of $\Omega$ satisfying these constraints, 
we first examine the case where
$\cos\Omega_{0}>0$ (the poles are in the first and fourth quadrants).
In this case $\zeta_z > 0$ and the upper limit  of (\ref{eq:range}) is at least 1.
Consequently, the inequality is always satisfied. This means $\cos\Omega$
needs to be in the range
\begin{equation}
\zeta_z-1 <\cos\Omega< 1
\end{equation}
and $\Omega$ in the range 
\begin{equation}
0 <\Omega< \cos^{-1}\left[\zeta_z-1\right]\label{eq:rangew1-1}
\end{equation}
Next we examine the case where $\cos\Omega_{0}<0$ (the poles are
in the second and third quadrants). In this case $\zeta_z < 0$ and the lower limit
 of (\ref{eq:range}) is $< -1$; hence, the inequality
is always satisfied. This means $\cos\Omega$ would be in the range
\begin{equation}
-1 <\cos\Omega< \zeta_z+1
\end{equation}
and $\Omega$ in the range 
\begin{equation}
\cos^{-1}\left[\zeta_z+1\right] <\Omega< \pi\label{eq:rangew2-1}
\end{equation}
 Last, for the case where $\cos\Omega_{0}=0$ (the poles are on the
imaginary axis), the inequality in (\ref{eq:range}) becomes
\begin{equation}
-1 <\cos\Omega< 1
\end{equation}
meaning that $\Omega$ is in the range 
$0 <\Omega< \pi$.
For the range to exist, we need the argument of the inverse cosine
in (\ref{eq:rangew1-1}) and (\ref{eq:rangew2-1}) to be less than
$1$ in magnitude. Starting with (\ref{eq:rangew1-1}), we can write
$-1 <\zeta_z-1< 1$
which yields, for $\cos\Omega_{0}>0$,
\begin{equation}
0 <\zeta_z< 2
\end{equation}
Similarly, starting with (\ref{eq:rangew2-1}), for
$\cos\Omega_{0}<0$ we have
\begin{equation}
-2 <\zeta_z< 0
\end{equation}
Combining the two cases along with the fact that $\cos\Omega_{0}=0$ also
produces a resonance yields the discrete-time damping constraint, Eq. 
\eqref{damping-z-constraint}.
Plugging in the definition of $\zeta_z$ and rearranging yields
\begin{equation}
A^{2}-\frac{2}{\left|\cos\Omega_{0}\right|}A+1 < 0
\end{equation}
We can factor the lefthand side and rewrite the inequality as 
\begin{equation}
\left(A-\frac{1-\left|\sin\Omega_{0}\right|}{\left|\cos\Omega_{0}\right|}\right)\left(A-\frac{1+\left|\sin\Omega_{0}\right|}{\left|\cos\Omega_{0}\right|}\right) < 0
\end{equation}
For the inequality to be satisfied, $A$ has to lie between the two
roots (to produce one negative term and one positive term), {\it i.e.},
\begin{equation}
\frac{1-\left|\sin\Omega_{0}\right|}{\left|\cos\Omega_{0}\right|}< A <\frac{1+\left|\sin\Omega_{0}\right|}{\left|\cos\Omega_{0}\right|}
\end{equation}
The lower bound describes a boundary line inside the unit circle while
the upper bound describes a boundary line outside the unit circle.
Pairs of poles in the region between these boundaries will produce
local maxima (resonances) in the frequency response. For causal, stable systems,
the region is defined by 
\begin{equation}
\frac{1-\left|\sin\Omega_{0}\right|}{\left|\cos\Omega_{0}\right|}< A < 1
\end{equation}
Hence, the boundary inside the unit circle 
is given by the curve 
\begin{equation}
z = \frac{1-\left|\sin\Omega_{0}\right|}{\left|\cos\Omega_{0}\right|}e^{j\Omega_{0}}\label{eq:boundary-1}
\end{equation}

 \end{document}